\documentclass[english,aps,pre,preprint,groupedaddress,showpacs]{revtex4-1}
\usepackage[T1]{fontenc}
\usepackage[latin9]{inputenc}
\usepackage{geometry}
\usepackage{float}
\usepackage{amsmath}
\usepackage{amssymb}
\usepackage{graphicx}
\usepackage{esint}

\makeatletter

\newcommand{\noun}[1]{\textsc{#1}}

\usepackage{mathrsfs}
\usepackage[english]{babel}

\makeatother

\usepackage{babel}
\begin{document}

\title{Long velocity tails in plasmas and gravitational systems}

\author{L.~Brenig\textsuperscript{1{*}}, Y.~Chaffi\textsuperscript{1}}

\affiliation{1. Physique des systèmes dynamiques. Faculté des Sciences. Université
Libre de Bruxelles (ULB). 1050 Brussels, Belgium.}

\author{T.~M.~Rocha Filho\textsuperscript{2}}

\affiliation{2. Instituto de Fí{}sica and International Center for Condensed
Matter Physics.\\
 Universidade de Brasí{}lia, CP: 04455, 70919-970 - Brasí{}lia,
Brazil.}

\email{lbrenig@ulb.ac.be}

\begin{abstract}
Long tails in the velocity distribution are observed in plasmas and
gravitational systems. Some experiments and observations in far-from-equilibrium
conditions show that these tails behave as $1/v^{5/2}$. We show here
that such heavy tails are due to a universal mechanism related to
the fluctuations of the total force field. Owing to the divergence
in $1/r^{2}$ of the binary interaction force, these fluctuations
can be very large and their probability density exhibits a similar
long tail. They induce large velocity fluctuations leading to the
$1/v^{5/2}$ tail. We extract the mechanism causing these properties
from the BBGKY hierarchy representation of Statistical Mechanics.
This leads to a modification of the Vlasov equation by an additional
term. The novel term involves a fractional power $3/4$ of the Laplacian
in velocity space and a fractional iterated time integral. Solving
the new kinetic equation for a uniform system, we retrieve the observed
$1/v^{5/2}$ tail for the velocity distribution. These results are
confirmed by molecular dynamics simulations.
\end{abstract}
\maketitle

\section{Introduction}

Long tails, i.e. asymptotic behaviour in $1/v^{\alpha}$, $\alpha>0$,
for large velocity $v$ of the velocity distribution are frequently
observed or inferred in experiments with far-from-equilibrium plasmas.
Similar tails are also obtained from astronomical data about large-scale
galaxy systems. More precisely, in experiments with focused ion beams~\cite{key-1},
the observed transverse velocity distribution of the ions is a symmetric
Lévy-stable distribution of stability index $3/2$~\cite{key-2}
with a long tail in $1/v^{5/2}$ for any component of the transverse
velocity vector. Other processes implying long tails in the velocity
distribution are the phenomena of ionisation and nuclear fusion in
plasmas. These processes are very sensitive to the number of high
velocity particles in the system. In these experiments the measured
rates are often significantly larger than those predicted using a
Gaussian velocity distribution~\cite{key-3}. Such discrepancies
led some researchers to replace the Gaussian velocity distribution
by a convolution between a Gaussian and a symmetric Lévy of index
$3/2$~\cite{key-4} . The rates calculated with this new distribution
came closer to the measured ones~\cite{key-3}.

\medskip{}

A qualitative explanation is the following. Consider a large system
of particles interacting via the Coulomb or the gravitational force
and in which spatial correlations between particles can be neglected.
The distribution of the random total force field $\vec{F}$ due to
all the particles is a Lévy-$3/2$ distribution, also called Holtsmark
distribution~\cite{key-5}~\cite{key-6}. The reduced Holtsmark
distribution for any component $F_{i}$ of the 3-dimensional vector
$\vec{F}$ has a long tail in $1/F_{i}^{5/2}$. This tail denotes
a much greater probability for large force fluctuations than with
a Gaussian distribution. It originates from the divergence at short
distances of the $1/r^{2}$ interaction force. Yet, for short enough
times, the velocity of a particle submitted to the total force $\vec{F}$
is essentially proportional to that field. Hence, for such short times,
the distribution of the velocities should be a convolution of a Lévy-$3/2$
distribution with the initial velocity distribution. For initial distributions
with finite second order moments, this convolution gives a heavy tail
in $1/v^{5/2}$~\cite{key-7-1}.

Similar results are obtained in large-scale systems of galaxies for
the distribution of the peculiar velocities of the galaxies~\cite{key-7}.
The peculiar velocity is the difference between the observed velocity
and the local cosmological expansion velocity. The observed peculiar
velocity distribution has a long tail in $1/v^{2.1}$. The discrepancy
between the exponent $2.1$ of the observed velocity tail and the
exponent $2.5$ of the tail of a true Holtsmark distribution is explained
by the fractal structure of the spatial distribution~\cite{key-8}~\cite{key-9}.

The Vlasov equation on which the current theories of plasmas and gravitational
systems mainly rely involves only the effects of the average total
force field. Since it excludes fluctuations around that average, this
equation cannot have Lévy distributions as generic solutions. A theoretical
explanation is thus required. In the present work we derive from the
BBGKY (Bogoliubov-Born-Green-Kirkwood-Yvon) hierarchy of equations~\cite{key-10}
a modified kinetic equation for systems interacting through forces
in $1/r^{2}$. It differs from the Vlasov equation by a new additive
contribution. That term involves a fractional power $3/4$ of the
Laplacian operator in velocity space and a fractional iterated integral
in time. The kinetic equation is derived in chapter 1. In chapter
2, its solution is obtained for a spatially uniform system and is
shown to possess the observed $1/v^{5/2}$ tail. In chapter 3, we
present molecular dynamics simulations that confirm this result. Conclusions
and perspectives are discussed in the last chapter.

\section{Deriving the kinetic equation from the BBGKY hierarchy}

We consider a system of $N$ identical classical point-like particles
of mass $m$ in $\mathbb{R}^{3}$. They interact via a binary potential
$U(\vec{r})=\gamma/r$. The variable $r$ is the norm of the distance
vector $\vec{r}$ between the two interacting particles. In order
to cover both repulsive and attractive interactions in charged particles
gases and gravitational systems, the coupling constant $\gamma$ can
be positive or negative.

In classical Statistical Mechanics a system is described by the 1-particle
phase-space distribution function (1-pdf) $f_{1}(\vec{r_{1}},\vec{v_{1};}t)\equiv f(\bold{1};t)$
and by the phase-space correlation functions of growing orders such
as $g_{2}(\bold{1},\bold{2};t)\equiv f_{2}(\bold{1},\bold{2};t)-f(\bold{1};t)f(\bold{2};t)$
where $f_{2}(\bold{1},\bold{2};t)$ is the 2-particles phase-space
distribution, $g_{3}$ and so on~\cite{key-10}. Here, the index
$\mathbf{\mathbf{i}=1,2,...,N}$ denotes the set of position and velocity
variables, $\vec{r}_{i}$, $\vec{v}_{i}$, of particle $\mathit{i}$
in the system. The time-evolution of the 1-pdf $f$ and the phase-space
correlations $g_{2}$, $g_{3}$, ... obey the so-called BBGKY hierarchy
of coupled equations~\cite{key-10}. Our aim is to derive from that
hierarchy a kinetic equation, that is, a closed equation for the 1-pdf
of the system. This can only be achieved by a truncation of the hierarchy.
This truncation must, of course, be based on some specific properties
of the system. In this chapter we present the main steps leading to
the kinetic equation for plasmas and gravitational systems. More details
are given in the Methods A section.

\medskip{}

The first equation of the BBGKY hierarchy~\cite{key-10} reads,

\begin{equation}
\partial_{t}f(\bold{1};t)=L_{1}^{0}f(\bold{1};t)+\int d\bold{2}\,L'_{12}\,f(\bold{1};t)\,f(\bold{2};t)+\,\int d\bold{2}\,L'_{12}\,g_{2}(\bold{1},\bold{2};t)\label{eq:1}
\end{equation}
The free motion operator $L_{i}^{0}$ corresponds to $L_{i}^{0}\equiv-\vec{v}_{i}\cdot\frac{\partial}{\partial\vec{r_{i}}}$.
The interaction operator $L_{ij}'$ is given by $L_{ij}'\equiv-\frac{1}{m}\vec{F}(\vec{r_{i}}-\vec{r}_{j})\cdot(\frac{\partial}{\partial\vec{v}_{i}}-\frac{\partial}{\partial\vec{v}_{j}})$
where $\vec{F}(\vec{r_{i}}-\vec{r}_{j})\equiv\vec{F}_{ij}\equiv\gamma\frac{\vec{r_{i}}-\vec{r}_{j}}{^{^{\left\Vert r_{i}-r_{j}\right\Vert ^{3}}}}$
is the interaction force of particle $\mathbf{\mathbf{\mathit{i}}}$
acting on $\mathbf{\mathit{j}}$. The integral symbol $\int d\boldsymbol{\mathbf{\mathbf{\mathbf{i}}}}$
stands for $\int d^{3}{r_{i}}\int d^{3}{v_{i}}$ where both integrals
are over $\mathbb{R}^{3}$. The integration domain over $\vec{r}_{i}$
should be the volume $V$ of the system, but in view of the thermodynamic
limit, $N\rightarrow\infty$, $V\rightarrow\infty$, $N/V=n=constant<\infty$,
considered here, the domain is assimilated to $\mathbb{\mathbb{R}^{\mathrm{3}}}$.
The thermodynamic limit should not be confused with the fluid limit
in which $N\rightarrow\infty$, $m\rightarrow0$, $\gamma\rightarrow0,$
$Nm=constant<\infty$, $N\gamma=constant<\infty$. The fluid limit
removes the discrete character of the particles. Since we are interested
in a phenomenon related to the discreteness of particles, the thermodynamic
limit is taken here. For a discussion of the two limits see reference~\cite{key-11-1}.

\medskip{}

In the right-hand-side of equation~(\ref{eq:1}), the first term
represents free motion while the second term denotes the Vlasov term.
The latter describes the effect of the mean force field due to all
the other particles on particle $\mathbf{1}$~\cite{key-10}. This
field is the average of the force over the 1-pdf itself. It vanishes
for uniform systems (see discussion in Methods A.1 after equation~(\ref{eq:63})).
The third term in the right-hand-side of equation~(\ref{eq:1}) takes
into account the 2-particles phase-space correlations $g_{2}$ and
couples this equation to the rest of the BBGKY hierarchy as we see
below.\medskip{}

Our concern is limited to weakly coupled systems, i.e. systems for
which $\Gamma\equiv U/kT\ll1$. Here, $kT$ denotes the average kinetic
energy of two particles. The quantity $U\equiv\left|U(\delta)\right|=\left|\gamma\right|/\delta$
represents the potential energy between two particles at the average
distance $\delta=n^{-1/3}$ between nearest neighbors. The weak coupling
condition is, thus, $\Gamma=\left|\gamma\right|n^{1/3}/kT\ll1$. We
also limit our scope to short time evolutions such that $t\ll t_{r}$
where $t_{r}$ is the relaxation time to equilibrium. For weakly coupled
systems, in the current theories, the third term $C\equiv\int d\bold{2}\,L'_{12}\,g_{2}(\bold{1},\bold{2};t)$
in equation~(\ref{eq:1}) is shown to contain the effect of binary
collisions~\cite{key-10}. Under these collisions the system irreversibly
relaxes towards equilibrium in a time $t_{r}\sim t_{s}\Gamma^{-3/2}$~\cite{key-11}
, where $t_{s}=(\frac{m}{4\pi\left|\gamma\right|n})^{1/2}$. The short
time-scale $t_{s}$ represents for plasmas, the plasma oscillations
period and for gravitational systems, the free fall time. Since $\Gamma\mathit{\ll\mathrm{1}}$,
$t_{r}$ is a very large time. Hence, for times $t\ll t_{r}$ such
as we consider, the term $C$ can be neglected. This reduces equation~(\ref{eq:1})
to the well-known Vlasov equation.

\medskip{}

However, for interaction forces that diverge as $1/r^{2}$ at small
distances the above reasoning does not hold. Statistically, the divergence
increases the probability of large fluctuations of the force which,
in turn, results in the long tail of the total force distribution.
This effect should be found in $C$ as the integral over $\vec{r}_{2}$
that it involves contains the near vicinity of the origin where the
divergence of the force occurs. That part of $C$, thus, cannot a
priori be neglected. To explicit this remark, we divide the integration
domain of the integral over $\vec{r_{2}}$ in $C$ in two parts: a
small open ball $S_{1}$ of radius $d$ centered at particle $\mathbf{1}$,
and the rest of the space, $\mathbb{R}\mathrm{^{3}\mathrm{\backslash\mathrm{\mathit{S}}_{1}}}$.
The radius $d$ is defined such that the average interaction energy
between any particle $2$ located in that sphere and particle $\bold{1}$
is larger than the sum of the average kinetic energies of these two
particles. We, thus, have $\frac{\left|\gamma\right|/d}{kT}$$=1$
which amounts to $d=\left|\gamma\right|/kT$. We also assume that
the typical macroscopic inhomogeneity length, $L_{H}$, is much larger
than $d$, $d\ll L_{H}$ and also that $\delta\ll L_{H}$. Combined
with $\Gamma\ll1$, this yields $d\ll\delta\ll L_{H}$.

We, thus, get

\begin{equation}
C=I_{1}+I_{2}\label{eq:2}
\end{equation}

with,

\begin{equation}
I_{1}=\int_{S_{1}}d^{3}r_{2}\int d^{3}v_{2}\,L'_{12}\,g_{2}(\vec{r}_{1},\vec{v}_{1},\vec{r}_{2},\vec{v}_{2};t)\label{eq:3}
\end{equation}

and,

\begin{equation}
I_{2}=\int_{\mathbb{R\mathrm{^{3}\mathrm{\backslash\mathrm{\mathit{S}}_{1}}}}}d^{3}r_{2}\int d^{3}v_{2}\,L'_{12}\,g_{2}(\vec{r}_{1},\vec{v}_{1},\vec{r}_{2},\vec{v}_{2};t)\label{eq:4}
\end{equation}

Owing to the above splitting, the norm of the interaction force $F_{12}\equiv\left\Vert \vec{F_{12}}\right\Vert $
is large in $I_{1}$ while it is small in $I_{2}$. As already mentioned,
the latter is known to lead to the collision term in the Landau or
the Balescu-Lenard equations~\cite{key-10}. Since our system is
assumed to be weakly coupled and, more importantly, as we are interested
in times $t\ll t_{r}$, the term $I_{2}$ can safely be neglected.
However, in view of what has been discussed above, by no means can
we neglect $I_{1}$ as is usually done. Equation~(\ref{eq:1}), hence,
becomes 
\begin{eqnarray}
\partial_{t}f(\bold{1};t) & = & L_{1}^{0}f(\bold{1};t)+\int d\bold{2}\,L'_{12}\,f(\bold{1};t)\,f(\bold{2};t)\label{eq:60-1}\\
 &  & +\int_{S_{1}}d^{3}r_{2}\int d^{3}v_{2}\,L'_{12}\,g_{2}(\vec{r}_{1},\vec{v}_{1},\vec{r}_{2},\vec{v}_{2};t)\nonumber 
\end{eqnarray}

Let us stress that in the standard derivations of the kinetic equations~\cite{key-10}
the discussion of $I_{1}$ is eluded. It is simply replaced by a cut-off
at short distances in the integral over $r_{2}$ in $C$ in order
to avoid that divergence. Usually, this cut-off is justified by assuming
an effective quantum repulsion at short distances. This means that
in these theories $C$, in fact, reduces to $I_{2}$. Consequently,
in these theories, for short time $t\ll t_{r}$ during which the effects
of collisions are negligible, the only remaining terms are the free
motion and the Vlasov mean-field terms. In other words, one gets the
Vlasov equation. In contrast, in the sequel we show that $I_{1}$
does not diverge (see Methods A.2 after equation~(\ref{eq:18-1}))
and the Vlasov equation must be modified by this term. We now proceed
to calculate $I_{1}$.

\medskip{}

A glance at equation~(\ref{eq:60-1}) shows that it is not closed.
It is coupled to the second equation of the BBGKY hierarchy~\cite{key-10}
via the unknown function $g_{2}(\vec{r}_{1},\vec{v}_{1};\vec{r}_{2},\vec{v}_{2};t)$.
Equation~(\ref{eq:60-1}) is, thus, coupled to the second equation
of the BBGKY hierarchy~\cite{key-10} 
\begin{eqnarray}
\partial_{t}g_{2}(\bold{1},\bold{2};t) & = & \big[{\textstyle L_{1}^{0}+L_{2}^{0}\big]g_{2}(\bold{1},\bold{2};t)+L'_{12}\big[g_{2}(\bold{1},\bold{2};t)+f(\bold{1};t)\,f(\bold{2};t)\big]}\nonumber \\
 &  & +\int d\bold{3}\{L'_{13}f(\bold{1};t)g_{2}(\bold{2},\bold{3};t)+L'_{23}f(\bold{2};t)g_{2}(\bold{1},\bold{3};t)\nonumber \\
 &  & +(L'_{13}+L'_{23})[f(\bold{3};t)g_{2}(\bold{1},\bold{2};t)+g_{3}(\bold{1},\bold{2},\bold{3},t)]\}\label{eq:6}
\end{eqnarray}
which must be considered with the constraint $\left\Vert \vec{r}_{2}-\vec{r}_{1}\right\Vert <d$.

Obviously, this equation is also not closed. It is coupled to the
third BBGKY equation -not written here- via the 3-particles phase-space
correlation $g_{3}$. The latter is coupled to the equation for $g_{4}$
and so on, generating the whole BBGKY hierarchy of coupled equations~\cite{key-10}.
However, as shown in Methods A.1, the conditions $\left\Vert \vec{r}_{2}-\vec{r}_{1}\right\Vert <d$,
$t\ll t_{r}$ along with $d\ll\delta\ll L_{H}$ allow for truncating
and greatly simplifying this hierarchy. Indeed, it turns out that
equation~(\ref{eq:6}) reduces to the closed equation,

\begin{equation}
\partial_{t}g_{2}(\bold{1},\bold{2};t)=L'_{12}\big[g_{2}(\bold{1},\bold{2};t)+f(\bold{1};t)\,f(\bold{2};t)\big]\label{eq:2.7}
\end{equation}

As shown in Methods A.2, the solution of this equation is easily found
and is introduced in the expression~(\ref{eq:3}) of $I_{1}$, yielding,

\begin{equation}
I_{1}\approx-\frac{1}{5}\left(\frac{2\pi\gamma}{m}\right)^{3/2}\int_{0}^{t}\frac{d\tau}{\sqrt{\tau}}\,\,n(\vec{r}_{1};t-\tau)\left(-\triangle_{\vec{v}_{1}}\right)^{3/4}\,\,f(\vec{r}_{1},\vec{v}_{1};t-\tau)\label{eq:8}
\end{equation}

where $n(\vec{r}_{1};t)\equiv\int d^{3}{v}f(\vec{r_{1}},\vec{v};t)$
is the local number density. The fractional power $3/4$ of the Laplacian
operator in the velocity variable is defined by 
\begin{equation}
\left(-\triangle_{\vec{v}_{1}}\right)^{3/4}e^{i\vec{\zeta}_{1}\cdot\vec{v}_{1}}\equiv\vec{(\zeta}_{1}.\vec{\zeta}_{1})^{^{3/4}}e^{i\vec{\zeta}_{1}\cdot\vec{v}_{1}}=\zeta_{1}^{3/2}e^{i\vec{\zeta}_{1}\cdot\vec{v}_{1}}\label{eq:14-2}
\end{equation}

and by using the Fourier integral representation of $f(\vec{r}_{1},\vec{v}_{1};t-\tau)$
with respect to the velocity.

Finally, replacing expression~(\ref{eq:8}) into equation~(\ref{eq:60-1}),
a closed equation for the 1-pdf is obtained. This is the final form
of our kinetic equation

\begin{align}
\mbox{}\partial_{t}f(\vec{r_{1},}\vec{v_{1}};t) & =-\vec{v}_{1}\cdot\vec{\nabla}_{1}f(\vec{r_{1}},\vec{v_{1};}t)\label{eq:15-2}\\
 & -\frac{1}{m}\int d^{3}{r_{2}}\int d^{3}{v_{2}}\,\vec{F}_{12}\cdot(\frac{\partial}{\partial\vec{v}_{1}}-\frac{\partial}{\partial\vec{v}_{2}})\,f(\vec{r_{1}},\vec{v_{1}};t)\,f(\vec{r_{2}},\vec{v_{2}};t)\mbox{\ensuremath{\mbox{}}}\nonumber \\
 & -\frac{1}{5}\left(\frac{2\pi\gamma}{m}\right)^{3/2}\int_{0}^{t}\frac{d\tau}{\sqrt{\tau}}\,\,n(\vec{r}_{1};t-\tau)\left(-\triangle_{\vec{v}_{1}}\right)^{3/4}\,\,f(\vec{r}_{1},\vec{v}_{1};t-\tau)\nonumber \\
\nonumber 
\end{align}

Equation~(\ref{eq:15-2}) is a Vlasov equation modified by the addition
of a new term. Like the Vlasov term, the new contribution is nonlinear
in the 1-pdf due to the presence in it of $n(\vec{r}_{1};t)$. Notice
that in solving equation~(\ref{eq:2.7}) the initial correlation
has been chosen to vanish (see Methods A.2). As discussed in chapter
5, a non-vanishing initial correlation would not change our main conclusions.

We next show that, in the case of homogeneous systems, the general
solution to the kinetic equation is a Lévy-$3/2$ distribution with
a long tail in $1/v^{5/2}$.

\section{Homogeneous systems}

We now consider the particular situation of a spatially uniform system.
For such systems the free motion term as well as the Vlasov term exactly
vanish in the kinetic equation~(\ref{eq:15-2}). This leaves only
the new term in that equation and singles out its effect on the evolution
of the 1-pdf.

For uniform systems the 1-pdf becomes $f(\vec{r},\vec{v,}t)=n\varphi(\vec{v,}t)$,
where $\varphi(\vec{v},t)$ is the velocity distribution at time $t$.
The equation~(\ref{eq:15-2}) then becomes 
\begin{equation}
\partial_{t}\varphi(\vec{v};t)=\,\,-\frac{n}{5}\left(\frac{2\pi\gamma}{m}\right)^{3/2}\int_{0}^{t}\frac{d\tau}{\sqrt{\tau}}\,\,\left(-\triangle_{v}\right)^{3/4}\,\,\varphi(\vec{v};t-\tau)\,\,\label{eq:22}
\end{equation}

Obviously, the kinetic equation becomes exactly linear. This offers
the possibility to find the exact solution of this equation. Such
is not the case for the general nonlinear equation~(\ref{eq:15-2}).
The solution of equation~(\ref{eq:22}) is easily found (see Methods
B). For short times, it reduces to a velocity-convolution between
the initial velocity distribution and a Lévy-$3/2$ distribution~\cite{key-2},
\begin{equation}
\varphi(\vec{v};t)\simeq\int d^{3}{u}\thinspace\thinspace\varphi(\vec{u};0)\thinspace\thinspace L_{3/2}(\vec{v}-\vec{u},\thinspace Ct^{3/2})\label{eq:52}
\end{equation}

with $C=\frac{4n}{15}\left(\frac{2\pi\gamma}{m}\right)^{3/2}$ and
where $L_{3/2}(\vec{v},\thinspace Ct^{3/2})$ denotes the probability
density of a multivariate isotropic Lévy-stable random variable $\vec{v}$
centered at zero with, in the notations of J.P.Nolan~\cite{key-2},
stability index $\alpha=3/2$ and scale factor $\tilde{\boldsymbol{\mathsf{\mathrm{\mathcal{\mathbf{\boldsymbol{\mathbb{\textrm{\ensuremath{\gamma}}}}}}}}}}=C^{2/3}t$.

For all initial distributions with finite second moments, the above
convolution gives a long-tailed distribution with tail $1/\mathsf{v}^{5/2}$~\cite{key-7-1}
where $\mathsf{v}$ is any component of the velocity vector $\vec{v}$.
This result alone justifies the derivation of the new kinetic equation~(\ref{eq:15-2})
as it establishes a clear connection with experimental and numerical
results as we see in next chapter.

\medskip{}

At this level, one could question the physical soundness of such a
distribution. Indeed, its second moment and all its higher order moments
diverge. Consequently, the average kinetic energy obtained from it
diverges. This divergence is due to the fact that particles that are
very near each other acquire very large accelerations and velocities
under the action of the divergent interaction force. However, this
never occurs in reality as natural cut-offs appear at very short distances.
For elementary particles they originate from quantum effective repulsion
and, for macroscopic bodies, from internal cohesion forces that maintain
their shape. This results in a form of regularisation of the potential
(see next chapter) and leads to a natural truncation of the tails
of these distributions. That truncation ensures the existence of the
second and higher order moments. Nevertheless, the long tail property
essentially persists, modified at only extremely large values of the
velocity by a sharp decrease.

\medskip{}

A last result is worth to be reported. Equation~(\ref{eq:22}) can
be cast into a particularly elegant form. The fractional iterated
integral operator of order $1/2$ acting on a function $f(t)$ is
defined~\cite{key-13} by $J_{t}^{1/2}f(t)\equiv\frac{1}{\sqrt{\pi}}\int_{0}^{t}\frac{d\tau}{\sqrt{\tau}}f(t-\tau)$.
Up to a factor $1/\sqrt{\pi}$ this is just the integral operator
on variable $\tau$ appearing in the right hand side of equation~(\ref{eq:22}).
The Riemann-Liouville fractional derivative of order $1/2$ in the
time variable $t$ is defined~\cite{key-13} as: $D_{t}^{1/2}\equiv\frac{d}{dt}\circ J_{t}^{1/2}$.
Let us apply $D_{t}^{1/2}$ on both sides of equation~(\ref{eq:22}).
Using the following properties and definitions~\cite{key-13}, $J_{t}^{1/2}\circ J_{t}^{1/2}=J_{t}^{1}$,
$\frac{d}{dt}\circ J_{t}^{1}=I$, $\frac{d}{dt}\circ D_{t}^{1/2}\equiv D_{t}^{3/2}$,
where $I$ is the identity operator, the equation~(\ref{eq:22})
transforms into 
\begin{equation}
D_{t}^{3/2}\varphi(\vec{v};t)=-A(-\triangle_{\vec{v}_{1}})^{3/4}\,\,\varphi(\vec{v};t)\,\,\label{eq:53}
\end{equation}

where $A\equiv\frac{3}{4}\pi^{1/2}\thinspace C$. To our knowledge,
this is the first time such a purely fractional partial differential
equation is derived from the basic principles of Statistical Mechanics.

\section{Numerical simulations}

We simulated a 3D gravitational system of 131,072 identical point-like
classical particles using a fourth order symplectic integrator. The
molecular dynamical code was implemented in the CUDA parallel computing
architecture on graphic processing units~\cite{key-12}. The initial
distribution of particles is spatially uniform in a spherical volume
with all the particles at rest and without space boundaries: The system
is open and particles can escape. The interaction potential $\frac{\gamma}{r}$
is regularized into $\frac{\gamma}{(r^{2}+\varepsilon^{2})^{1/2}}$
in order to avoid divergences in the numerical integration. The statistical
significance is increased by performing 100 realisations (runs). 

\medskip{}

\begin{figure}[H]
\includegraphics[scale=0.3]{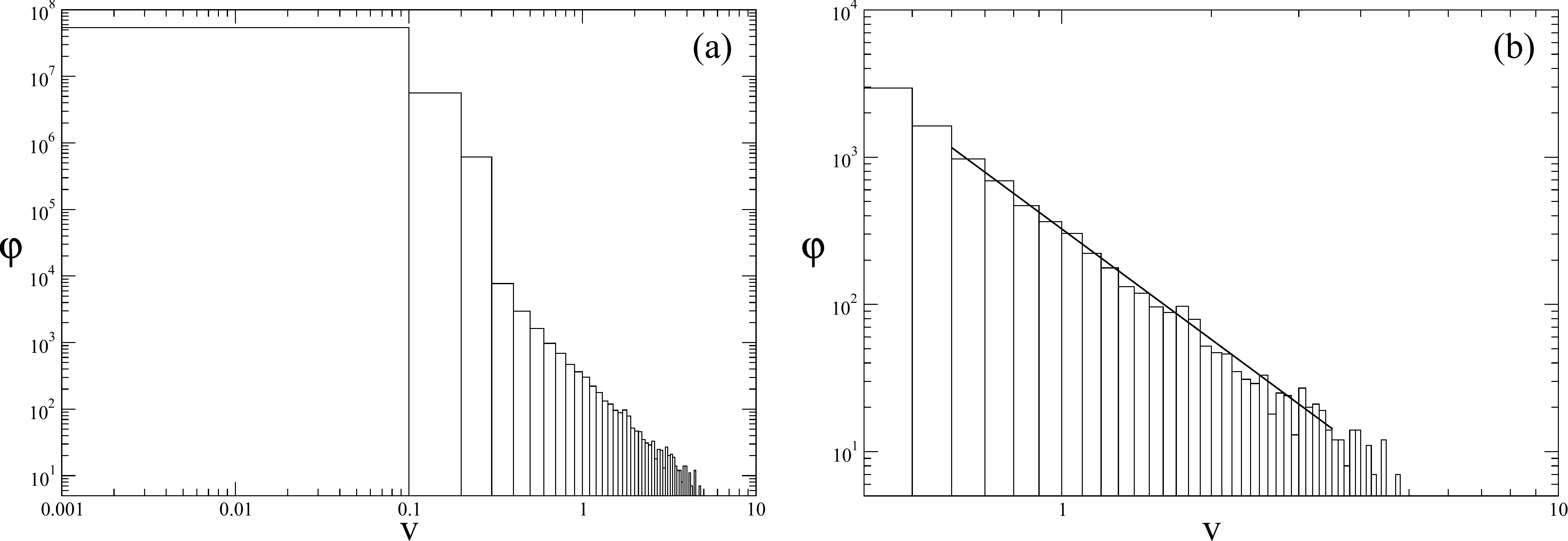}

\caption{(a) Log-log histogram of distribution of any component $\mathsf{v}$
of velocity $\vec{v}$ for regularisation parameter $\varepsilon=10^{-16}$.
Total number of particles $N=131,072$. Number of realisations $=100$.
Initial velocities $=0.00$. Initial spatial distribution: uniform
in a sphere of radius $R=1.28$. Time step $=2\times10^{-7}T$, total
run time $=3\times10^{-6}T$ with $T=(nGm)^{-1/2}$, $G$ is the gravitational
constant. \protect \\
(b) Zoom on the tail of the velocity distribution given in (a). The
thick straight line is the result of a linear regression on the tail
with slope $=-2.49$, standard error $=0.13$ and correlation coefficient
$=-0.973$.}

\label{fig1}
\end{figure}

After a very short time, long tails proportional to $1/\mathsf{v}^{\alpha}$
develop in the distribution for any component $\mathsf{v}$ of the
velocity vector $\vec{v}$ (see\ Figures~\ref{fig1}\ a\ and~\ref{fig1}b).

\begin{figure}[h]
\includegraphics[scale=0.5]{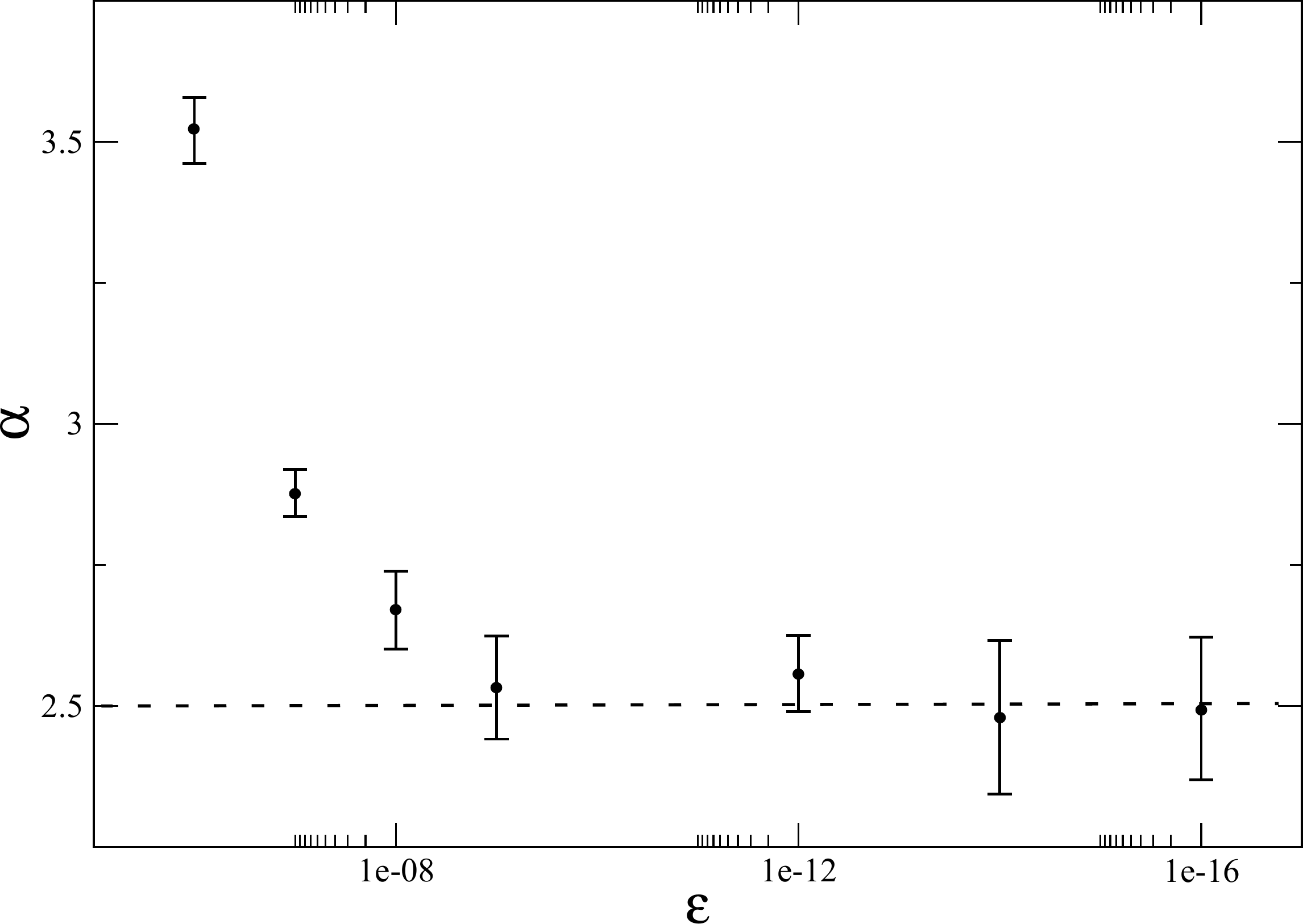}

\caption{Semi-log graph of the tail exponent $\alpha$ in $1/\mathsf{v}^{\alpha}$
where $\mathsf{v}$ is one component of velocity $\vec{v}$, as a
function of the regularisation parameter $\varepsilon$. Each point
is obtained from a linear regression on velocity data obtained from
100 realisations (runs) as in Figure 1.b. Error bars correspond to
the standard deviation obtained from the linear regression and the
dashed line is the theoretical value at $\alpha=2.5$.}

\label{fig2}
\end{figure}

Figure~\ref{fig2} shows the exponent $\alpha$ for decreasing values
of $\varepsilon$ and same initial conditions. Each point corresponds
to 100 runs (realisations). As observed, $\alpha$ approaches the
theoretical value $5/2$ as $\varepsilon$ gets smaller. 

The tail of the distribution is very sensitive to the behavior of
the interaction force at small distances and, consequently, to $\varepsilon$,
such that the expected behaviour is observed only for very small values
of $\varepsilon$. Note that the errors (from $5\%$ to $10\%$) on
$\alpha$ should not appear as a great concern as they could be greatly
reduced by increasing either the number of particles in the simulation
or the number of realisations used for statistical significance. Indeed,
the tail of a distribution is much more sensitive to the finiteness
of $N$ than its bulk. The reason is that the $1/\mathsf{v}^{\alpha}$
tail of a Lévy distribution is a consequence of a generalized Central-Limit
theorem~\cite{key-2} which only holds at the limit $N$$\rightarrow\infty$.
Yet, numerical simulations (and real systems too) are bound to involve
a finite number of particles. As a consequence, the number of particles
with very large values of the velocity, i.e. those that constitute
the tail of the distribution, is finite and represents only a small
fraction of $N$. The shape of the tail, thence, is subject to large
fluctuations.

\section{Perspectives and conclusions}

The new term in the kinetic equation~(\ref{eq:15-2}) is of order
$1/N$ compared to the Vlasov term. However, in uniform or near-uniform
spatial configurations, while the Vlasov term vanishes or is very
small, this term remains finite and, consequently, cannot be neglected.
In fully inhomogeneous states, furthermore, though small compared
to the Vlasov term, this new term might have some important consequences.
Indeed, different finite samples of $N$ points obeying a statistical
distribution with an algebraic tail usually have very different standard
deviations (diverging with $N$). As a consequence, this new term
may have a measurable influence on the quasi-stationary state that
appears after the violent relaxation process~\cite{key-14}~\cite{key-15}.
Also, its magnitude is of the same order as the collisional corrections
to the Vlasov equation. The latter become important for long times.
Therefore, an extension of the present approach to the relaxation
time-scale may be of interest. Longer simulation runs are required
to study how the long tails disappear and how the new term affects
the evolution of the system on longer time scales.

\medskip{}

In our derivation of the kinetic equation~(\ref{eq:15-2}) we assumed
a vanishing initial binary correlation function (see Methods A.2).
What would be the effect of a non-vanishing initial correlation? We
studied this question and, briefly, the results are the following.
Non-vanishing initial correlations introduce a source term in the
kinetic equation. The solution of that equation involves a time convolution
between this source term and the propagator of equation~(\ref{eq:15-2}).
Using a theorem in reference~\cite{key-7-1}, one then shows for
uniform systems that the tail of the resulting distribution remains
proportional to $1/\mathsf{v}^{5/2}$ for a large class of initial
correlations.

\medskip{}

The present theory extends without difficulty to systems with more
general interaction forces that behave as $1/r^{2}$ only at short
distances as, for instance, systems interacting via a Yukawa-type
or Debye-screened potential $\gamma\frac{e^{-r/\lambda}}{r}$ . Moreover,
if in addition one has $\lambda\ll L_{H}$, i.e. the interaction is
short-ranged, then the Vlasov term is negligeable~\cite{key-10}
and the supplementary term we derived is dominant in the kinetic equation.
Yukawa-like effective potentials also play an important role in nuclear
physics and, more particularly, in heavy-ion collisions such as those
occurring in the large accelerators. However, the nuclear effective
interaction is more complex than the Yukawa potential\cite{key-16}.
The former depends on the spins and isospins of the two interacting
particles. In some spin and isospin states the potential diverges
as $1/r$ at short distances, in others it behaves as $1/r^{3}$.
Quantum effects are, thus, important in heavy-ion interactions and
would require a quantum or, at least, a semi-classical extension of
our approach.

\section{Methods}

\subsection*{A. Derivation of the kinetic equation~(\ref{eq:15-2}) }

\subsubsection*{1. Truncation of the BBGKY hierarchy for $\left\Vert \vec{r}_{2}-\vec{r}_{1}\right\Vert <d$}

Let us analyse the integral term in equation~(\ref{eq:6}). We denote
it by $K$ 
\begin{eqnarray}
K & =\int_{\mathbb{R}^{3}}d^{3}{r_{3}}\int_{\mathbb{R}^{3}}d^{3}{v_{3}}\Big\{ L'_{13}f(\bold{1};t)g_{2}(\bold{2},\bold{3};t)+L'_{23}f(\bold{2};t)g_{2}(\bold{1},\bold{3};t)+\nonumber \\
 & (L'_{13}+L'_{23})\big[f(\bold{3};t)g_{2}(\bold{1},\bold{2};t)+g_{3}(\bold{1},\bold{2},\bold{3};t)\big]\Big\}\\
\nonumber 
\end{eqnarray}

$K$ is the sum of four contributions $K=K_{1}+K_{2}+K_{3}+K_{4}$
that are defined below, one after the other. The first one is 
\begin{equation}
K_{1}=\int_{\mathbb{R}^{3}}d^{3}{r_{3}}\int_{\mathbb{R}^{3}}d^{3}{v_{3}}L'_{13}\,f(\bold{1};t)g_{2}(\bold{2},\bold{3};t)\label{eq:55}
\end{equation}

or more explicitly and with a permutation of integrals 
\begin{equation}
K_{1}=-\frac{1}{m}\int_{\mathbb{R}^{3}}d^{3}{v_{3}}\int_{\mathbb{R}^{3}}d^{3}{r_{3}}\vec{F}(\vec{r_{1}}-\vec{r}_{3}).(\frac{\partial}{\partial\vec{v}_{1}}-\frac{\partial}{\partial\vec{v}_{3}})f(\vec{r}_{1},\vec{v}_{1};t)g_{2}(\vec{r}_{2},\vec{v}_{2},\vec{r}_{3},\vec{v}_{3};t)\label{eq:56}
\end{equation}

where $\vec{F}(\vec{r})\equiv\gamma\frac{\vec{r}}{r^{3}}$ . The part
of the volume integral over $\vec{v}_{3}$ containing $\frac{\partial}{\partial\vec{v}_{3}}$
transforms into a surface integral on the surface at infinity in the
sub-space of velocity $\vec{v}_{3}$ and vanishes due to the fact
that $g_{2}(\vec{r}_{2},\vec{v}_{2},\vec{r}_{3},\vec{v}_{3};t)\rightarrow0$
for $v_{3}\rightarrow\infty$~\cite{key-10}. Let us make successively
two changes of variable in the integral over $\vec{r}_{3}$: First,
$\vec{r}_{3}\rightarrow\vec{r}=\vec{(r_{1}}-\vec{r}_{3})$ and, second,
$\vec{r}\rightarrow\vec{F}=\gamma\frac{\vec{r}}{r^{3}}$ . In the
last transformation the volume element becomes $d^{3}r=\frac{1}{2}\gamma^{3/2}F^{-9/2}d^{3}F$.
Hence, $K_{1}$ reads now 
\begin{equation}
K_{1}=-\frac{\gamma^{3/2}}{2m}\int_{\mathbb{R}^{3}}d^{3}{v_{3}}\int_{\mathbb{R}^{3}}d^{3}{F}\,F^{-9/2}\,\vec{F}\thinspace g_{2}(\vec{r}_{2},\vec{v}_{2},\vec{r}_{1}-\gamma^{1/2}F^{-3/2}\vec{F},\vec{v}_{3};t).\frac{\partial}{\partial\vec{v_{1}}}f(\vec{r}_{1},\vec{v}_{1};t)\label{eq:57}
\end{equation}

We now express the integral over $\vec{F}$ in spherical coordinates
$F$, $\theta$, $\varphi$ 
\begin{multline}
\begin{aligned}\end{aligned}
K_{1}=-\frac{\gamma^{3/2}}{2m}\int_{\mathbb{R}^{3}}d^{3}{v_{3}}\int_{0}^{\pi}d\theta\thinspace sin\theta\int_{0}^{2\pi}d\varphi\thinspace\vec{n}(\theta,\varphi)\\
\int_{0}^{\infty}dF\thinspace F^{-3/2}g_{2}(\vec{r}_{2},\vec{v}_{2},\vec{r}_{1}-\gamma^{1/2}F^{-1/2}\vec{n}(\theta,\varphi),\vec{v}_{3};t).\frac{\partial}{\partial\vec{v_{1}}}f(\vec{r}_{1},\vec{v}_{1};t)\label{eq:58}
\end{multline}

where $\vec{n}(\theta,\varphi)$ is the unit vector 
\[
\vec{n}(\theta,\varphi)=\left(\begin{array}{c}
sin\theta\thinspace cos\varphi\\
sin\theta\thinspace sin\varphi\\
cos\theta
\end{array}\right)
\]

Finally, the change of variable $F\rightarrow u=F^{-1/2}$ yields
\begin{align}
\begin{aligned}\end{aligned}
K_{1} & =-\frac{\gamma^{3/2}}{m}\int_{\mathbb{R}^{3}}d^{3}{v_{3}}\int_{0}^{\pi}d\theta\thinspace sin\theta\int_{0}^{2\pi}d\varphi\thinspace\vec{n}(\theta,\varphi)\label{eq:59}\\
 & \int_{0}^{\infty}du\thinspace g_{2}(\vec{r}_{2},\vec{v}_{2},\vec{r}_{1}-\gamma^{1/2}u\thinspace\vec{n}(\theta,\varphi),\vec{v}_{3};t).\frac{\partial}{\partial\vec{v_{1}}}f(\vec{r}_{1},\vec{v}_{1};t)\nonumber 
\end{align}

Notice that the divergence of the integral over $\vec{r_{3}}$ in
equation~(\ref{eq:56}) that could have been expected from the divergence
of the force when $\vec{r}_{3}\rightarrow\vec{r}_{1}$ does not occur
here. Indeed, in its transformed form~(\ref{eq:59}) the integral
over $u$ contains only $g_{2}$ which, in turn, must be an integrable
function of all its arguments. The last claim comes from the fact
that the two-particles phase-space distribution must be integrable
in order to be normalised. Hence, for all values of $\vec{r_{1}}$
and $\vec{r_{2}}$ the term $K_{1}$ is finite. This contrasts with
the term $\mathcal{L\equiv}L'_{12}\big[g_{2}(\bold{1},\bold{2};t)+f(\bold{1};t)\,f(\bold{2};t)\big]$
of equation~(\ref{eq:6}) where in $\mathcal{\mathit{L'_{\mathsf{12}}}}$
the force diverges for $\vec{r}_{2}\rightarrow\vec{r}_{1}$. Since
equation~(\ref{eq:6}) is considered here with the constraint $\left\Vert \vec{r}_{2}-\vec{r}_{1}\right\Vert <d$,
$\mathcal{L}$ is dominant over $K_{1}$. The same argument applies
to $K_{2}$ with the permutation $1\longleftrightarrow2$ 
\begin{equation}
K_{2}=\int_{\mathbb{R}^{3}}d^{3}{r_{3}}\int_{\mathbb{R}^{3}}d^{3}{v_{3}}L_{23}\,f(\bold{2};t)g_{2}(\bold{1},\bold{3};t)
\end{equation}

.

Using the same changes of variables as above, the term $K_{3}$ 
\begin{equation}
K_{3}=\int_{\mathbb{R}^{3}}d^{3}{r_{3}}\int_{\mathbb{R}^{3}}d^{3}{v_{3}}(L'_{13}+L'_{23})g_{3}(\bold{1},\bold{2},\bold{3};t)\label{eq:60}
\end{equation}

becomes 
\begin{align}
K_{3} & =-\frac{\gamma^{3/2}}{m}\frac{\partial}{\partial\vec{v}_{1}}.\int_{\mathbb{R}^{3}}d^{3}{v_{3}}\int_{0}^{\pi}d\theta\thinspace sin\theta\int_{0}^{2\pi}d\varphi\thinspace\vec{n}(\theta,\varphi)\int_{0}^{\infty}du\thinspace g_{3}(\vec{r}_{1},\vec{v}_{1,}\vec{r}_{2},\vec{v}_{2},\vec{r}_{1}-\gamma^{1/2}u\thinspace\vec{n}(\theta,\varphi),\vec{v}_{3};t)\nonumber \\
 & -\frac{\gamma^{3/2}}{m}\frac{\partial}{\partial\vec{v}_{2}}.\int_{\mathbb{R}^{3}}d^{3}{v_{3}}\int_{0}^{\pi}d\theta\thinspace sin\theta\int_{0}^{2\pi}d\varphi\thinspace\vec{n}(\theta,\varphi)\int_{0}^{\infty}du\thinspace g_{3}(\vec{r}_{1},\vec{v}_{1,}\vec{r}_{2},\vec{v}_{2},\vec{r}_{2}-\gamma^{1/2}u\thinspace\vec{n}(\theta,\varphi),\vec{v}_{3};t)
\end{align}

and with a similar argument as for $K_{1}$ and $K_{2}$ one can neglect
$K_{3}$ with respect to $\mathcal{L}$. Finally, let us consider
the term $K_{4}$ 
\begin{equation}
K_{4}=\int_{\mathbb{R}^{3}}d^{3}{r_{3}}\int_{\mathbb{R}^{3}}d^{3}{v_{3}}(L'_{13}+L'_{23})f(\bold{3};t)g_{2}(\bold{1},\bold{2};t)\label{eq:62}
\end{equation}

More explicitly 
\begin{equation}
K_{4}=-\frac{1}{m}\{\int_{\mathbb{R}^{3}}d^{3}{r_{3}}\vec{F}(\vec{r_{1}}-\vec{r}_{3})n(\vec{r}_{3};t).\frac{\partial}{\partial\vec{v}_{1}}+\int d^{3}{r_{3}}\vec{F}(\vec{r_{2}}-\vec{r}_{3})n(\vec{r}_{3};t).\frac{\partial}{\partial\vec{v}_{2}}\}g_{2}(\bold{1},\bold{2};t)\label{eq:63}
\end{equation}

where $n(\vec{r};t)$ is the local number density defined in Chapter
2. The integral $\mathcal{\vec{F}}(\vec{r}_{1})\equiv\int d^{3}{r_{3}}\vec{\,F}(\vec{r_{1}}-\vec{r}_{3})\,n(\vec{r}_{3};t)$,
the Vlasov mean force field, represents $N$ times the statistical
mean of the force that another particle 3 exerts on particle 1 averaged
on the position probability density $p(\vec{r_{3}};t)\equiv\frac{n(\vec{r_{3}};t)}{N}$.
The second integral has the same meaning but with particle 1 replaced
by particle 2. Let us, then, compare $K_{4}$ to the term $\mathcal{L}$
written more explicitly as 
\begin{equation}
\mathcal{L}=-\frac{1}{m}\vec{F}(\vec{r_{1}}-\vec{r}_{2}).(\frac{\partial}{\partial\vec{v}_{1}}-\frac{\partial}{\partial\vec{v}_{2}})[g(\bold{1},\bold{2};t)+f(\bold{1};t)\,f(\bold{2};t)]\label{eq:64}
\end{equation}

We must compare the orders of magnitude of $\left\Vert \vec{F}(\vec{r_{1}}-\vec{r}_{2})\right\Vert $
and $\left\Vert \mathcal{\vec{F}}(\vec{r}_{i})\right\Vert $, $i=1,2$,
for $\left\Vert \vec{r}_{2}-\vec{r}_{1}\right\Vert <d$. One has $\left\Vert \vec{F}(\vec{r_{1}}-\vec{r}_{2})\right\Vert >\frac{\gamma}{d^{2}}$
. As to $\mathcal{\vec{F}\mathrm{(\mathit{\vec{r}_{i}\mathrm{)}}}}$,
using integration by part, it can be rewritten as $\mathcal{\vec{F}}(\vec{r}_{i})\equiv-\int d^{3}{r_{3}}\,U(\vec{r_{i}}-\vec{r}_{3})\,\frac{\partial}{\partial\vec{r}_{3}}n(\vec{r}_{3};t)$
where we used the fact that the potential $U(\vec{r})\rightarrow0$
for $r\rightarrow\infty$ and $n(\vec{r,}t)\rightarrow n=N/V$ for
$r\rightarrow\infty$. Clearly, $\mathcal{\vec{F}}(\vec{r}_{1})$
vanishes for homogeneous systems. The integrand in $\mathcal{\vec{F}}(\vec{r}_{i})$
is vanishingly small in every part of the integration domain where
the gradient of the local number density, $\left\Vert \frac{\partial}{\partial\vec{r}}n(\vec{r};t)\right\Vert $,
is vanishingly small. Let us call $L_{H}$ the typical length on which
$n(\vec{r};t)$ varies noticeably. Thus, the volume of integration
in which the integrand does not vanish is of order $L_{H}^{3}$. Consequently,
the order of magnitude of the integral defining $\left\Vert \mathcal{\vec{F}}(\vec{r}_{i})\right\Vert $
is $\frac{\gamma}{L_{H}}.\frac{n}{L_{H}}.L_{H}^{3}=\gamma nL_{H}$.
Hence, $K_{4}$ is negligible with respect to $\mathcal{L}$ if $\frac{\gamma}{d^{2}}>\gamma nL_{H}$.
This inequality transforms into $\mathit{\Gamma^{\mathrm{2}}<\mathrm{\frac{\delta}{\mathit{L_{H}}}\ll1}}$
which is compatible with the physical conditions formulated in Chapter
2.

The free motion term $\big[L_{1}^{0}+L_{2}^{0}\big]g_{2}(\bold{1},\bold{2};t)$
of equation~(\ref{eq:6}) is also negligible compared to the term
$\mathcal{L}$. The latter is proportional to the force between particles
1 and 2 which is of order $\Gamma^{-\mathrm{2}}$. Indeed, one has
$\left\Vert \vec{F}(\vec{r_{1}}-\vec{r}_{2})\right\Vert >\frac{\gamma}{d^{2}}=\mathit{\Gamma^{-\mathrm{2}}\gamma n^{\mathrm{2/3}}}$
while the free motion operator $\big[L_{1}^{0}+L_{2}^{0}\big]$ is
independent of $\mathit{\Gamma}$.

With these arguments, what remains from equation~(\ref{eq:6}) is
equation~(\ref{eq:2.7}).

\subsubsection*{2. Establishing the kinetic equation~(\ref{eq:15-2})}

The equation~(\ref{eq:2.7}) is solved by adding the solution of
the homogeneous part of this equation to the convolution of the propagator
of the homogeneous equation with the source term $L'_{12}f(\bold{1};t)\,f(\bold{2};t)$.
Using the Fourier-transform with respect to the velocities and some
simple algebra, one gets 
\begin{align}
\tilde{g}_{2}(\vec{r}_{1},\vec{\zeta}_{1},\vec{r}_{2},\vec{\zeta}_{2};t) & =\tilde{U}(t)\,\tilde{g}_{2}(\vec{r}_{1},\vec{\zeta}_{1},\vec{r}_{2},\vec{\zeta}_{2};0)\nonumber \\
 & +\frac{\partial}{\partial\alpha}\int_{0}^{t}\frac{d\tau}{\tau}\,\,\tilde{U}(\alpha\tau)\,\,\tilde{f}(\vec{r}_{1},\vec{\zeta}_{1};t-\tau)\,\tilde{f}(\vec{r}_{2},\vec{\zeta}_{2};t-\tau)\mid_{\alpha=1}\label{eq:13-1}
\end{align}
where $\vec{\zeta_{1}}$ and $\vec{\zeta_{2}}$ are the Fourier variables
associated to the velocities $\vec{v_{1}}$ and $\vec{v_{2}}$ and
where 
\begin{equation}
\tilde{U}(t)=\exp[(-\frac{i}{m}\vec{F}_{12}\cdot(\vec{\zeta}_{1}-\vec{\zeta}_{2}))t]
\end{equation}

is the Fourier-transform of the propagator of the homogeneous part
of equation~(\ref{eq:2.7}).

From here on, vanishing initial correlation $g_{2}(\vec{r}_{1},\vec{v}_{1};\vec{r}_{2},\vec{v}_{2};0)$
is assumed. As discussed in chapter 5, non-vanishing initial correlation
would not change our main conclusions.

\medskip{}

Taking the inverse Fourier transform of expression~(\ref{eq:13-1}),
introducing the resulting formula for $g_{2}(\bold{1},\bold{2};t)$
in the formula~(\ref{eq:3}) of $I_{1}$ and after a permutation
of integrals, one obtains 
\begin{align}
I_{1} & =\,\,\frac{\partial}{\partial\alpha}\int d^{3}v_{2}\int\frac{d^{3}\zeta_{1}d^{3}\zeta_{2}}{(2\pi)^{6}}\,\,e^{i\vec{\zeta}_{1}\cdot\vec{v}_{1}+i\vec{\zeta}_{2}\cdot\vec{v}_{2}}\nonumber \\
 & \int_{0}^{t}\frac{d\tau}{\tau}\int_{S_{1}}d^{3}r_{2}\,\frac{(-i)}{m}\vec{F}_{12}\cdot(\vec{\zeta}_{1}-\vec{\zeta}_{2})\,e^{-\frac{i\alpha}{m}\vec{F}_{12}\cdot(\vec{\zeta}_{1}-\vec{\zeta}_{2})\tau}\tilde{f}(\vec{r}_{1},\vec{\zeta}_{1};t-\tau)\,\tilde{f}(\vec{r}_{2},\vec{\zeta}_{2};t-\tau)\mid_{\alpha=1}\label{eq:17.1}
\end{align}

Since $d\ll L_{H}$, one can approximate $\tilde{f}(\vec{r}_{2},\vec{\zeta}_{2};t-\tau)$
by its value at $\vec{r_{2}}=\vec{r}_{1}$ and extract it from the
integral over $\vec{r_{2}}$ in the ball $S_{1}$. With some algebra,
equation~(\ref{eq:17.1}) becomes 
\begin{equation}
I_{1}\approx\,\,-\frac{\partial^{2}}{\partial\alpha^{2}}\int\frac{d^{3}\zeta_{1}}{(2\pi)^{3}}\,\,e^{i\vec{\zeta}_{1}\cdot\vec{v}_{1}}\int_{0}^{t}\frac{d\tau}{\tau^{2}}\,\,\tilde{f}(\vec{r}_{1},\vec{\zeta}_{1};t-\tau)\,\,n(\vec{r}_{1};t-\tau)\thinspace\thinspace J\mid_{\alpha=1}\label{eq:15-1}
\end{equation}

with 
\begin{equation}
J\equiv\int_{\mathcal{\mathcal{\mathrm{\mathit{S_{\mathrm{1}}}}}}}d^{3}r\,\,\,\left(e^{-\frac{i\alpha}{m}\vec{F}(\vec{r})\cdot\vec{\zeta}_{1}\tau}-1\right)\label{eq:51}
\end{equation}

After a change of variable $\vec{r}\rightarrow\vec{F}\vec{(r)}$ and
passing to spherical coordinates, $J$ transforms into 
\begin{equation}
J=-2\pi\left(\frac{\gamma\zeta_{1}\alpha\tau}{m}\right)^{3/2}\left(\frac{2}{3}(z_{m})^{-3/2}-\int_{z_{m}}^{\infty}dz\,\,z^{-7/2}\sin z\right)\,\,\label{eq:50}
\end{equation}

where $z_{m}\equiv\frac{\gamma\alpha\tau\zeta_{1}}{d^{2}m}$. The
above integral is an incomplete Sine-integral function whose power-series
expansion in $z_{m}$ (see \cite{key-17}) leads to 
\begin{equation}
J=2\pi d^{3}\thinspace[\frac{4}{15}\sqrt{2\pi}\left(\frac{\gamma\alpha\tau\zeta_{1}}{d^{2}m}\right)^{3/2}-\frac{1}{3}\left(\frac{\gamma\alpha\tau\zeta_{1}}{d^{2}m}\right)^{2}+\frac{1}{300}\left(\frac{\gamma\alpha\tau\zeta_{1}}{d^{2}m}\right)^{4}+\frac{1}{3740}\left(\frac{\gamma\alpha\tau\zeta_{1}}{d^{2}m}\right)^{6}+\cdots]\label{eq:18-1}
\end{equation}

We, now, can discuss the question raised in Chapter 2 (after equation~(\ref{eq:60-1}))
about the convergence of the integral $I_{1}$. In its form~(\ref{eq:15-1}),
the only place where the force $\vec{F}(\vec{r})$ appears is the
integral $J$ given by equation~(\ref{eq:51}). As seen from its
result~(\ref{eq:18-1}), $J$ converges. This comes from the fact
that the force appears only in a phase factor in equation~(\ref{eq:51}).

\medskip{}

Coherently with our short-time assumption, we suppose $z_{m}\ll1$
and retain only the first term of the series. The upper boundary $t$
of the time integral in equation~(\ref{eq:15-1}), thus, must be
such that $\frac{\gamma\alpha t\zeta_{1}}{d^{2}m}\ll1$. More explicitly,
let us replace $\zeta_{1}$ by the inverse of an average velocity
$v_{av}$ and put $\alpha=1$. This transforms the previous inequality
into $\frac{\gamma}{d^{2}m}t\ll v_{av}$. In other words, the time
$t$ must be such that the velocity increment $\Delta v=\frac{\gamma}{d^{2}m}t$
acquired by particle $1$ during time $t$ under the force of another
particle at the surface of $S_{1}$, satisfies $\Delta v\ll v_{av}$.

Finally, introducing the first term of series~(\ref{eq:18-1}) in
equation~(\ref{eq:15-1}), one gets equation~(\ref{eq:8}) which,
in turn, leads to the kinetic equation~(\ref{eq:15-2}) in Chapter
2.

\subsection*{B. Solution of the equation~\ref{eq:22}}

A Fourier transform with respect to $\vec{v}$ and a Laplace transform
with respect to $t$ of equation~(\ref{eq:22}) give 
\begin{equation}
\hat{\tilde{\varphi}}(\vec{\zeta};w)=\frac{w^{1/2}\,\,\,\tilde{\varphi}(\vec{\zeta};0)}{w^{3/2}\,\,\,+\,\,\,A\,|\vec{\zeta}|^{3/2}}\label{eq:23-1}
\end{equation}

where $\tilde{\varphi}(\vec{\zeta};0)$ is the Fourier transform of
$\varphi(\vec{v},t)$ at $t=0$, $\hat{\tilde{\varphi}}(\vec{\zeta};w)$
is the Fourier-Laplace transform of $\varphi(\vec{v};t)$, and $A=\frac{n\sqrt{\pi}}{5}\left(\frac{2\pi\gamma}{m}\right)^{3/2}$.
The inverse Laplace transform of $\hat{\tilde{\varphi}}(\vec{\zeta};w)$
is taken by first expanding equation~(\ref{eq:23-1}) in powers of
$A\,|\vec{\zeta}|^{3/2}$ and, then, integrating the series term by
term. This leads to the exact solution of equation~(\ref{eq:22})

\begin{equation}
\varphi(\vec{v};t)=\int\frac{d^{3}\zeta}{(2\pi)^{3}}e^{i\vec{\zeta}\cdot\vec{v}}\,\,\tilde{\varphi}(\vec{\zeta};0)\,\,E_{3/2}(-A\,\,\zeta^{3/2}\,\,t^{3/2})\label{eq:24-1}
\end{equation}

where 
\begin{equation}
E_{\mu}(u)=\sum_{k=0}^{\infty}\frac{(u)^{k}}{\Gamma(\mu k+1)}
\end{equation}

is the Mittag-Leffler function of parameter $\mu$~\cite{key-13}.
The condition $z_{m}\ll1$ assumed in our derivation of the kinetic
equation is obviously compatible with condition $A\,\,\zeta^{3/2}\,\,t^{3/2}\ll1$.
We, thus, can safely make the following approximation

\begin{equation}
E_{3/2}(-A\,\zeta^{3/2}\,t^{3/2})\simeq e^{-C\,(\,\zeta\,t\,)^{3/2}}\label{eq:26-1}
\end{equation}

with $C=\frac{4n}{15}\left(\frac{2\pi\gamma}{m}\right)^{3/2}$. Finally,
we get 
\begin{equation}
\varphi(\vec{v};t)\simeq\int\frac{d^{3}\zeta}{(2\pi)^{3}}e^{i\vec{\zeta}\cdot\vec{v}}\,\,\tilde{\varphi}(\vec{\zeta};0)\,\,e^{-C\,(\,\zeta\,t\,)^{3/2}}
\end{equation}

equivalent to the velocity convolution of the initial velocity distribution
and a Lévy-$3/2$ distribution~\cite{key-2}

\begin{equation}
\varphi(\vec{v};t)\simeq\int d^{3}{u}\thinspace\thinspace\varphi(\vec{u};0)\thinspace\thinspace L_{3/2}(\vec{v}-\vec{u},\thinspace Ct^{3/2})
\end{equation}

Using a theorem in reference~\cite{key-7-1} one then shows that
for any $\varphi(\vec{v},0)$ with finite second moments, this approximation
as well as the exact solution~(\ref{eq:24-1}) have a long tail in
$1/\mathsf{v}^{5/2}$ where $\mathsf{v}$ is any component of the
velocity vector $\vec{v}$. 

\begin{acknowledgments}
The authors are greatly indebted to Drs. J.Wallenborn (ULB) and A.Figueiredo
(UnB) for the many deep and fruitful discussions during the elaboration
of the present work.\end{acknowledgments}

\end{document}